%% file: main.tex
\documentclass[english]{article}

\pdfoutput=1
\usepackage{lmodern}
\usepackage{booktabs}
\usepackage[T1]{fontenc}
\usepackage{subfloat}
\usepackage[latin9]{inputenc}
\usepackage{babel}

\makeatletter

\usepackage{mathrsfs}   
\usepackage{slashed}     
\usepackage{url}
\usepackage{graphicx}
\usepackage{color} 
\usepackage{bm}
\usepackage[numbers,sort&compress]{natbib}
\usepackage{float}
\usepackage{multirow}
\usepackage{amsmath}
\usepackage{amssymb}
\usepackage{breqn}
\usepackage{subcaption}
\usepackage{caption}
\usepackage{mciteplus}
\usepackage{authblk}
\usepackage{diagbox}
\usepackage{stackengine}
\usepackage[normalem]{ulem}

\input{Packages/myTikz}
\input{Packages/myHypref}

\newcommand{\T}{\mathrm{T}}
\newcommand{\kmax}{k^2_{\text{\footnotesize max}}}
\newcommand{\KaTie}{Ka\hspace{-0.2ex}Tie}

\begin{document}

\title{
\hspace{\fill}{\normalsize IFJPAN-IV-2024-1}\\[4ex]
\bf Examination of $k_t$-factorization in a Yukawa theory}
\author[1]{B. Guiot\thanks{benjamin.guiot@usm.cl}}
\author[2]{A. van Hameren}
\affil[1]{\it{\normalsize Departamento de F\'isica, Universidad T\'ecnica Federico Santa Mar\'ia; Casilla 110-V, Valparaiso, Chile}}
\affil[2]{\it{\normalsize Institute of Nuclear Physics Polisch Academy of Sciences,
PL-31342 Krak\'ow, Poland}}

\renewcommand\Authands{ and }
\date{}

\maketitle
\begin{abstract}
We discuss the upper limit, $k_{\text{\footnotesize max}}$, of the transverse-momentum integration performed in the $k_t$-factorization formula. Based on explicit calculations in the Yukawa theory and the study of seminal papers, we argue that $k_{\text{\footnotesize max}}$ is equal to the factorization scale $\mu_F$ used to factorize the cross section into an off-shell hard coefficient and a universal factor. There is consequently a relation between $k_{\text{\footnotesize max}}$ and the definition of unintegrated parton densities (UPDFs). The use of an inconsistent relation leads potentially to the overestimation of the cross section, which has been observed, e.g., in D-meson production \cite{Guiot2019}. One of our conclusions is that UPDFs related to collinear PDFs by an integration up to $\mu\sim Q$, where $Q$ is the hard scale and $\mu$ the scale in the collinear PDFs, imply that $\kmax\sim Q^2$. Integrating the transverse-momentum significantly above may result in the overestimation of the cross section. On the opposite, for UPDFs related to the collinear ones by an integration of the transverse momentum up to infinity, any $\kmax > Q^2$ is fine.
\end{abstract}
\newpage

\tableofcontents

\section{Motivation \label{secmotiv}}
In Ref.~\cite{Aslan2023}, the authors studied the accuracy of factorization for the deep-inelastic scattering (DIS) and semi-inclusive DIS structure functions. To do so, they worked with a scalar Yukawa theory given by
\begin{align}
    \mathcal{L}=&\sum_j\frac{i}{2}\left[\overline{\psi}_j\gamma^\mu \partial_\mu \psi_j-(\partial_\mu \overline{\psi}_j)\gamma^\mu \psi_j\right]-M_j\overline{\psi}_j\psi_j + \frac{1}{2}(\partial \phi)^2-\frac{m_s^2}{2}\phi^2\\
    &-\lambda\left(\overline{\psi}_1 \psi_2\phi + \overline{\psi}_2 \psi_1\phi \right), \nonumber
\end{align}
where $j=1,2$, $\psi_1=\psi_N$ represents the target ``nucleon'' with mass $M_1=m_p$, $\psi_2=\psi_q$ is the ``quark'' field with mass $m_q$, and $\phi$ is the ``scalar gluon''. The interest of this theory is that everything can be calculated perturbatively, including the exact (not factorized) DIS diagrams and parton densities.\\

\begin{figure}[!h]
\hspace*{-1.4cm}
 \includegraphics[width=35pc]{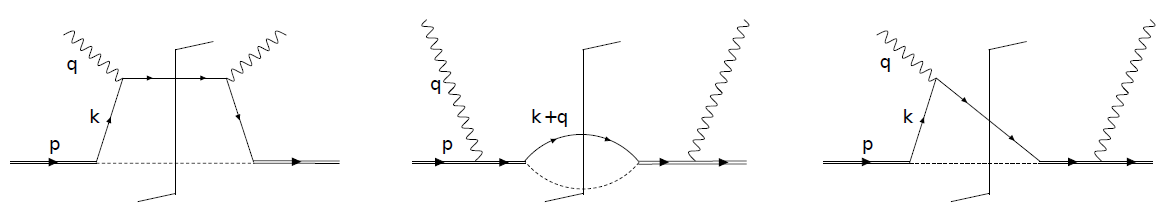}
\caption{\label{exact} Order $\mathcal{O}(\lambda^2)$ diagrams for the DIS on a quark target in the Yukawa theory.}
\end{figure}

A central equation for the present discussion is the factorization of the transverse-momentum-dependent (TMD) structure function \cite{Aslan2023}
\begin{align}
F_2\left(x_{\mathrm{bj}}, Q, \boldsymbol{k}_{\mathrm{T}}\right)= & \mathrm{T}_{\mathrm{small}} F_2\left(x_{\mathrm{bj}}, Q, \boldsymbol{k}_{\mathrm{T}}\right)+\mathrm{T}_{\mathrm{large}}\left[F_2\left(x_{\mathrm{bj}}, Q, \boldsymbol{k}_{\mathrm{T}}\right)\right. \nonumber\\
& \left.-\mathrm{T}_{\mathrm{small}} F_2\left(x_{\mathrm{bj}}, Q, \boldsymbol{k}_{\mathrm{T}}\right)\right]+\mathcal{O}\left(\frac{m^2}{Q^2}\right).\label{factF}
\end{align}
Here, $\boldsymbol{k}_\T$ is the transverse momentum of the initial quark, $Q$ is the usual DIS variable, $m$ generically refers to $m_p$, $m_q$, and $m_s$, and
\begin{equation}
    x_{\mathrm{bj}}=\frac{Q^2}{2P.q}
\end{equation}
is the Bjorken $x$. The variables $P$ and $q$ are shown on Fig.~\ref{exact}, and correspond to the target and photon 4-momentum.
We choose $m_p=m_q=m=0.5$ GeV and $m_s=1$ GeV. These values follow the hierarchy used in the appendix A4 of Ref.~\cite{Aslan2023}. $\mathrm{T}_{\mathrm{small}}$ and $\mathrm{T}_{\mathrm{large}}$ correspond to approximations in the computation of Feynman graphs and will be explained later. In our case, we want to study the $k_t$-factorization \cite{Catani1990,Catani1991,Collins1991,Levin91}, or high-energy factorization, in DIS and proton-proton collisions. With this formalism, reviewed in Sec.~\ref{secktfact}, the structure function reads
\begin{equation}
    F_2(x_{\mathrm{bj}},Q)=\int_{Q^2/s}^1dx\int_0^{\kmax}d^2k_{\T} F_q(x,k_{\T};Q)\hat{F}_2(x,Q,k_\T).\label{ktfact}
\end{equation}
 $F_q$ is the unintegrated parton distribution function (UPDF), here for a ``quark'' inside a ``nucleon'', and $\hat{F}_2$ is the gauge-invariant hard coefficient computed (usually) for an off-shell initial quark. Since the $k_t$-factorization was historically designed for small $x_{\mathrm{bj}}$, we will choose the Mandelstam variable $s\gg Q^2$, with values corresponding to the EIC kinematics. One of our main goals is to discuss the upper limit of integration, $\kmax$. Maybe due to the fact that the Balitsky-Fadin-Kuraev-Lipatov (BFKL) equation \cite{Fadin1975,Kuraev1976,Kuraev1977,Balitsky1978}, governing the cross sections and UPDFs evolution with $x$, has no ultraviolet divergence, the precise definition of $\kmax$ is generally not discussed. It is then not always possible to know if the results presented in phenomenological studies are obtained for $\kmax=\infty$, $s$, or $Q^2$, with $Q^2$ the hard scale. While, in several cases, these three choices lead to numerically close results, we want to discuss situations where a precise and consistent definition of $\kmax$ does matter. This discussion is related to an important (and generally ignored) issue, namely, the overestimation of some cross sections by $k_t$-factorization calculations, e.g. for open heavy-flavor production. \\

\begin{figure}[!h]
\begin{center}
 \includegraphics[width=26pc]{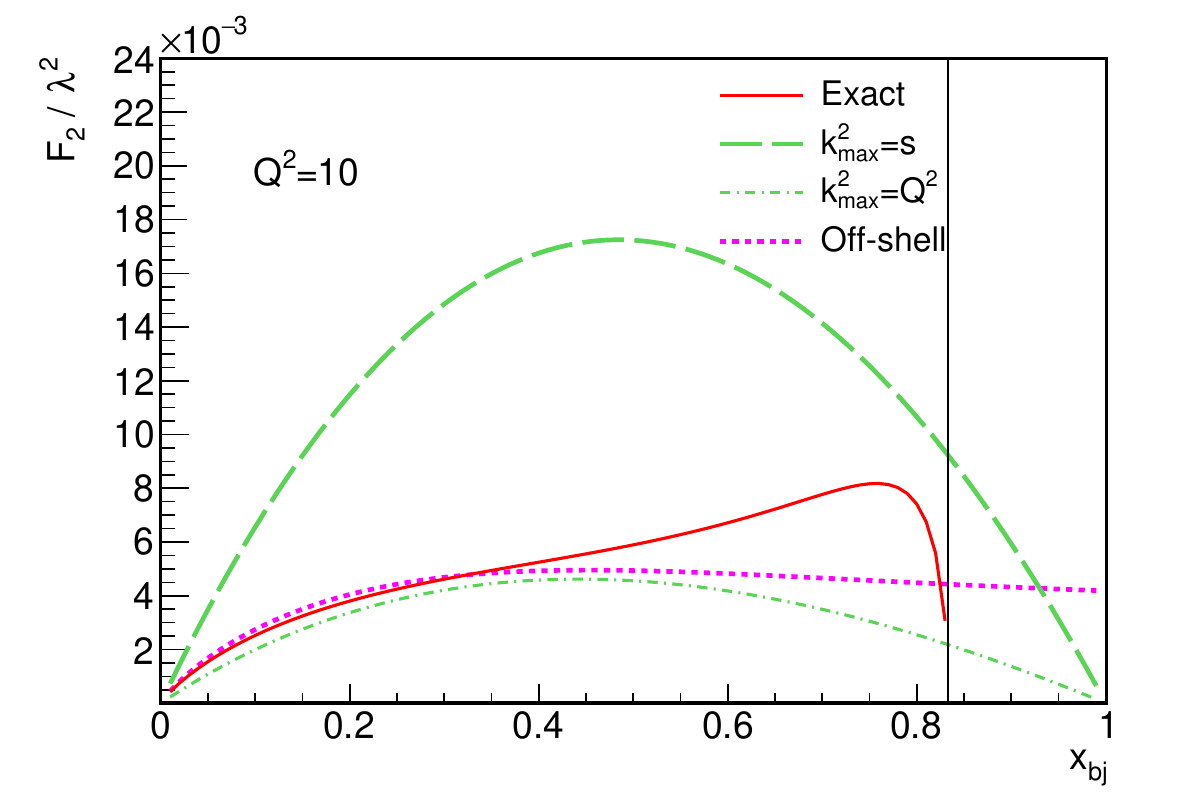}
\end{center}
\caption{\label{f2comp} Comparison of the exact result with $k_t$-factorization calculations for the $F_2$ structure function. The dashed line corresponds to Eq.~(\ref{ktfact}) with $\kmax=s$ and the hard coefficient on-shell. The dashed-dotted line is similar but with the integration limit changed to $Q^2$. The dotted pink line corresponds to Eq.~(\ref{ktfact}) with the off-shell hard coefficient given in Eq.~(\ref{f2offex}). The black vertical line indicates the kinematical limit for the exact case.}
\end{figure}

\begin{figure}[t!]
\begin{center}
 \includegraphics[width=22pc]{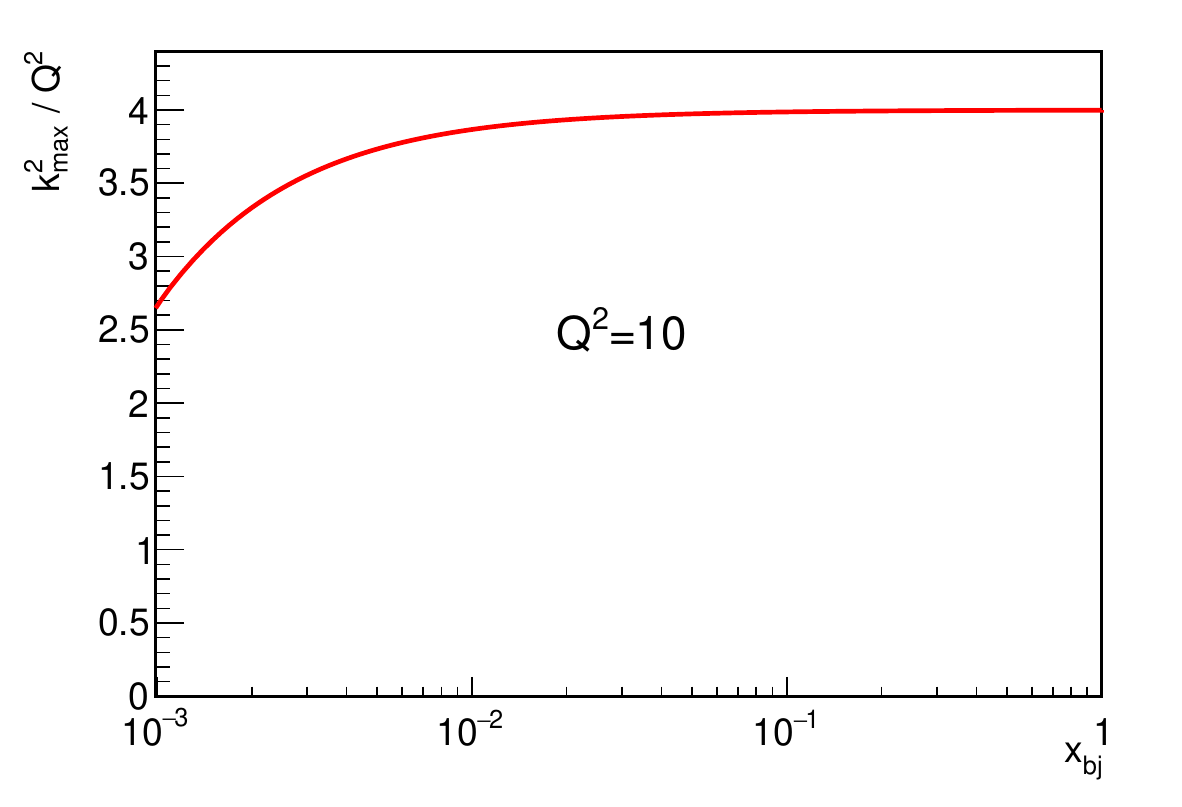}
\end{center}
\caption{\label{ratioktmQ2} Ratio of $\kmax$ given by Eqs.~(\ref{f2offex})-(\ref{kap}) to the DIS variable $Q^2$ as a function of $x_{\text{bj}}$. The convolution variable is fixed at $x=x_{\text{bj}}$.}
\end{figure}

To motivate our study, we compare the $k_t$-factorization and exact results at order $\mathcal{O}(\lambda^2)$ for the $F_2$ structure function, with $\lambda$ the coupling of the Yukawa theory. Here, exact means no factorization: The diagrams of Fig.~\ref{exact} can be calculated exactly, see appendix A4 of Ref.~\cite{Aslan2023}.
For the $k_t$-factorization calculations presented in Fig.~\ref{f2comp}, we used the Kimber-Martin-Ryskin-Watt (KMRW) UPDFs \cite{Kimber2001,Watt2003} computed in the Yukawa theory\footnote{See Refs.~ \cite{GolecBiernat2018,Hautmann2019,Guiot2020a,Valeshabadi2022,Nefedov2020,Guiot2021,Guiot2023} for discussions on these UPDFs in QCD.}, see Sec.~\ref{secKMR}, and, for now, an on-shell hard coefficient.
We choose the Mandelstam variable $s$ corresponding to an electron and proton beams of $30$ GeV and $250$ GeV, i.e., $s=3\times 10^4$ GeV$^2$. While the dashed green line obtained with Eq.~(\ref{ktfact}) and $\kmax=s$ overestimates the exact structure function, the same calculation with $\kmax=Q^2$ (dashed-dotted green line) gives a much better result. However, one could argue that the $k_t$-factorization should use off-shell coefficients. In the case of DIS, we have
\begin{equation}
     \hat{F}_2^{\text{ off-shell}}(x,Q,k_\T)=e_q^2\frac{\Theta(\Delta(x,k_\T))}{\sqrt{\Delta(x,k_\T)}}, \label{f2offex}
\end{equation}
with  
\begin{equation}
    \Delta(x,k_\T)=\left(\kappa_+-\frac{k_\T^2}{Q^2}\right)\left(\frac{k_\T^2}{Q^2}-\kappa_- \right),\label{deltaxk}
\end{equation}
and
\begin{equation}
    \kappa_\pm(x) = \left(\sqrt{1-y} \pm \sqrt{x/x_{\text{bj}}-y}\right)^2.\label{kap}
\end{equation}
Here, $y$ is defined by
\begin{equation}
    y=\frac{Q^2}{x_{\text{bj}} s}.
\end{equation}
Calculations of off-shell coefficients use the approximation $k^2=-k^2_\T$, related to the strong ordering in rapidity of the BFKL equation. Putting Eq.~(\ref{f2offex}) into Eq.~(\ref{ktfact}), we obtain the result shown as a dotted pink line in Fig.~\ref{f2comp}, in fair agreement with the exact result. Because of the off-shellness of $k$, $\kmax$ is now constrained by the theta function of Eq.~(\ref{f2offex}) and depends on $Q^2$, $x_{\text{bj}}$, and $x$, the convolution variable. Since the main contribution to the structure function comes from $x\sim x_{\text{bj}}$, we plot in Fig.~\ref{ratioktmQ2} the ratio $\kmax/Q^2$ as a function of $x=x_{\text{bj}}$.
We observe that $\kmax$ and $Q^2$ are of the same order of magnitude and conclude that in both cases, on-shell and off-shell, an acceptable result (at not too large $x_{\text{bj}}$) is obtained for $\kmax\simeq Q^2$. On the opposite, $\kmax =s$, when allowed by the kinematics, overestimates the exact result. We see that, at least in this specific case, the choice of $\kmax$ matters. The relevance of this first result stems from the fact that the issues encountered in the Yukawa theory are generally present in QCD.\\

Indeed, it is also possible to find examples in QCD, for instance, with the transverse-momentum distribution of heavy mesons in proton-proton collisions. Using the event generator \KaTie\ \cite{Hameren2018} and the KMRW UPDFs, Ref.~\cite{Guiot2019} showed that the full result (including flavor excitation and flavor creation contributions) overestimates the cross section by a factor of 4,\footnote{Note that the corresponding calculation in collinear factorization works perfectly fine, at any order in $\alpha_s$.} while the same calculation with the transverse-momentum integral cut at $k_\T=Q$ gave a good description of the data. It implies that in \KaTie , for this kinematical configuration, the integrals on the two initial partons' transverse momenta are performed well above $Q^2$, and the contribution from the region $k^2_\T>Q^2$ is not negligible. To avoid any discussions on flavor creation and excitation processes, as well as on hadronization, we can give simpler example. 
\begin{figure}[!t]
\begin{center}
 \includegraphics[width=20pc]{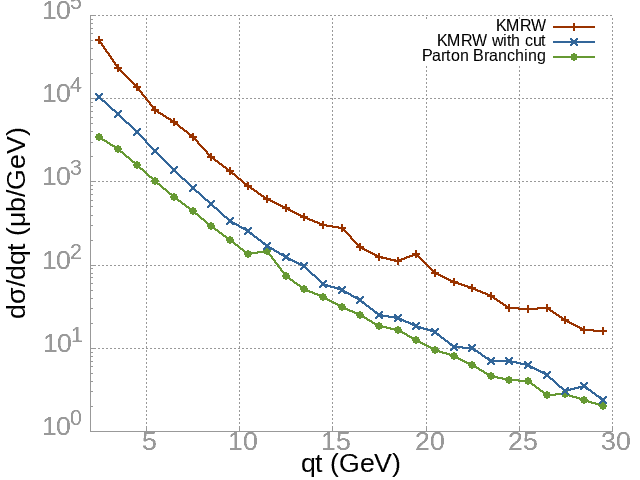}
\end{center}
\caption{\label{ggtogg}  Result obtained for the $gg\to gg$ cross section with the event generator \KaTie\ and different UPDFs. The variable $q_t$ is the gluon transverse momentum.}
\end{figure}
Using again the event generator \KaTie, we present in Fig.~\ref{ggtogg} the transverse-momentum distribution of one gluon after a $gg\to gg$ scattering. To avoid any kind of divergences, the transverse momentum of the two outgoing gluons is required to be larger than 2 GeV, and the gluon is ``measured'' in the rapidity range $|y|<1$. We set the scale $\mu$ according to the standard equation $\mu=(q_t+p_t)/2$, with $q_t$ and $p_t$ the transverse momentum of the measured and second gluon, respectively. We observe that the red line corresponding to standard KMRW UPDFs (with the angular ordering cutoff) is well above the results obtained with the Parton-Branching (PB) \cite{Hautmann2017,Hautmann2018,Martinez2019} and the ``cut KMRW''  (set equal to zero if $k_\T>Q$) UPDFs, demonstrating again the possibility of a non-negligible contribution of the integration above $Q^2$ in proton-proton collision. The fair agreement between the PB and cut KMRW UPDFs is not a coincidence, the value of the former being negligible in the region $k_\T>Q$.\\

From these few plots, we conclude that a discussion on $\kmax$ is mandatory.
After a quick remainder on the high-energy factorization in Sec.~\ref{secktfact}, and its comparison with the TMD factorization in Sec.~\ref{seccomp}, we show in Sec.~\ref{seckmax} that $k_{\text{\footnotesize max}}$ is the factorization scale $\mu_F$, and will explain its connection with the definition of UPDFs. In Sec.~\ref{secKMR}, we play with the KMRW UPDFs in the Yukawa theory and discuss a special kind of UPDF. Finally, we give our conclusion in Sec.~\ref{secQCD}.\\

\section{High-energy factorization\label{secktfact}}
In this section, we present various implementations of $k_t$, or high-energy, factorization. This formalism was developed in the 1990s by three distinct groups \cite{Catani1990,Catani1991,Collins1991,Levin91}. The primary focus was on addressing the energy dependence of the cross-section and the resummation of large $\ln s$, where $s$ represents the Mandelstam variable for the colliding system. This approach employed the BFKL equation, resumming the large logarithms of energy to all orders. In short, one starts with a cross section obeying the BFKL equation, and factorizes out a universal resummation factor (also solution of the BFKL equation), which is then combined to collinear PDFs to form the UPDFs
\begin{equation}
    F_N(k_\T;\mu)=C_N(k_\T,\alpha_s;\mu)f_N(\mu),
\end{equation}
In perturbative calculations, one generally chooses $\mu \simeq Q$ to limit the impact of higher orders. Quantities with the index $N$ are Mellin transforms of $x$-dependent functions, for instance,
\begin{equation}
    f_N(\mu)=\int_0^1 dx x^{N-1}f(x;\mu),
\end{equation}
for the collinear PDF. The resummation factor in Mellin space reads
\begin{equation}
    C_N(k_\T,\alpha_s;\mu)=R_N(\gamma_N(\alpha_s))\frac{\gamma_N(\alpha_s)}{k_\T^2}\left(\frac{k_\T^2}{\mu^2}\right)^{\gamma_N(\alpha_s)}, \label{resumfa}
\end{equation}
with the resummed anomalous dimension defined implicitly by
\begin{equation}
    \frac{\alpha_s C_A}{\pi}\chi(\gamma_N(\alpha_s))=N-1.
\end{equation}
$C_A$ is the usual color factor, and 
\begin{equation}
    \chi(\gamma)=2\psi_0(1)-\psi_0(\gamma)-\psi_0(1-\gamma),
\end{equation}
is the Lipatov's characteristic function, with $\psi_0$ is the digamma function. Off-shell cross sections are computed in the $k_t$-factorization scheme where the factor $R_N$ is equal to 1. Its value in the modified-minimal-subtraction scheme is given in \cite{Catani1993}. With $R_N=1$, the form of the resummation factor leads to the following relation between unintegrated and collinear PDFs
\begin{equation}
    \int_0^{\mu^2}dk_\T^2 F_N(k_\T;\mu) = f_N(\mu) \Rightarrow \int_0^{\mu^2}dk_\T^2 F(x,k_\T;\mu) = f(x;\mu)\label{UCN}
\end{equation}
Several comments are in order:
\begin{itemize}
    \item Relation~(\ref{UCN}) is not obeyed order by order in $\alpha_s$.\footnote{The perturbative expansion of the resummed anomalous dimension starts by $\frac{\alpha_sC_A}{\pi(N-1)}$, giving $F_N(k_\T;\mu)=\frac{\alpha_sC_A}{\pi(N-1)k_\T^2}f_N(\mu)+\mathcal{O}(\alpha_s^2)$.}
    \item UPDFs are convoluted with an off-shell hard coefficient, as shown in Eq~(\ref{ktfact}). This off-shellness is directly related to the BFKL equation and the factorization of the resummation factor.
    \item Eq.~(\ref{UCN}) is the consequence of using the BFKL equation to resum the large logarithms of $s$.
\end{itemize}

Another kind of $k_t$-factorization formalism starts from relation~(\ref{UCN}) (generally with a factor $x$ multiplying the collinear PDF) to build UPDFs. It is the case for the KMRW formalism discussed in Sec.~\ref{secKMR}. Here, the UPDFs do not obey the BFKL equation, and the focus is on transverse-momentum-dependent observables rather than the energy dependence of the cross section. Despite the UPDFs built in this formalism not obeying the BFKL equation, this version of $k_t$-factorization uses off-shell coefficients.\\

Finally, we can also mention an implementation of the high-energy factorization using UPDFs evolved according to the Ciafaloni-Catani-Fiorani-Marchesini (CCFM) and CCFM-Kwieci\'nski  equations \cite{Ciafaloni1988,Catani1990a,Gawron2004}. In principle, this version is close to the original one, since the CCFM and BFKL equations are equivalent in the high-energy limit \cite{Salam1999,Avsar2009}. Note that, at realistic energies, these two equations are not equivalent, and the quark contribution cannot be neglected.\\

Despite their differences, our discussion on $\kmax$ will apply to these implementations of high-energy factorization. We did not mention the parton-branching formalism \cite{Hautmann2017,Martinez2019}, which uses on-shell coefficients and shares similarities with both high-energy and TMD factorization. It will be useful for our discussion to compare these formalism, which is done in the next section.

\section{Comparison with the TMD factorization \label{seccomp}}
\begin{figure}[!h]
\begin{center}
 \includegraphics[width=8pc]{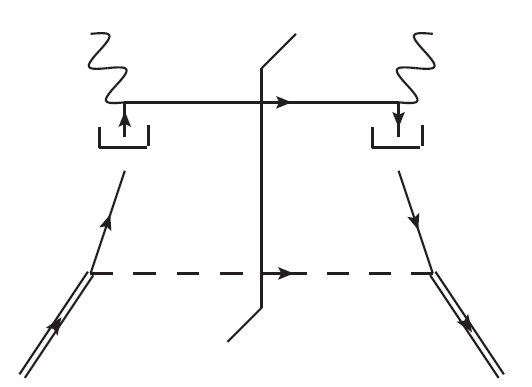}
\end{center}
\caption{\label{wterm} Diagram corresponding to the W-term. The lines above the hook should be treated as massless and on-shell.}
\end{figure}

There are two important differences between Eqs.~(\ref{factF}) and (\ref{ktfact}). The most visible is that Eq.~(\ref{ktfact}) corresponds to the first term in the r.h.s of Eq.~(\ref{factF}). Indeed, at order $\mathcal{O}(\lambda^2)$ we have
\begin{equation}
    \mathrm{T}_{\mathrm{small}} F_2 = \hat{F}_2^{(0)}\otimes f_{q/p}^{(1)},
\end{equation}
with $\hat{F}_2^{(0)}$ the LO hard coefficient and $f_{q/p}^{(1)}$ the TMD PDF at order $\mathcal{O}(\lambda^2)$. This expression corresponds to the Diagram shown in Fig.~\ref{wterm}.

\noindent On the opposite, the second term
\begin{equation}
    \mathrm{T}_{\mathrm{large}} F_2 = \hat{F}_2^{(1)}\otimes f_{q/p}^{(0)}
\end{equation}
and the subtraction term, $-\mathrm{T}_{\mathrm{large}}\mathrm{T}_{\mathrm{small}} F_2$, are not present in the high-energy factorization. The first term of Eq.~(\ref{factF}) is usually called the W-term, while the two others form the Y-term. Both depend on the factorization scale $\mu$ in such a way that the dependence cancels. It is worth mentioning that the Y- and W-terms are of the same order in $\lambda$.\\

A second difference between these two formalisms, related to the overestimation in Fig.~\ref{f2comp}, is the treatment of the W-term. While the TMD factorization integrates the transverse momentum up to infinity and includes a counterterm to renormalize the UV divergence, $k_t$-factorization uses a cutoff $\kmax$. As explained in \cite{Aslan2023}, at leading order, including the $\overline{\text{MS}}$ counterterm is equivalent to cut the integral at $k_\T^2=\mu^2$. Then, the second difference is that the TMD factorization systematically cut the integral at $\mu\sim Q$, independently of the details of the short-distance process, while $\kmax$ is frequently chosen much larger than the hard scale (as demonstrated in Sec.~\ref{secmotiv} by the discussion on gluon-gluon collision) and depends probably\footnote{The details on $\kmax$ being in general not provided, it is a guess. At least, it is true for our example in DIS with the off-shell coefficient Eqs.~(\ref{f2offex})-(\ref{kap})} on the hard-coefficient kinematics.\\

Note that for the $\mathcal{O}(\lambda^2)$ calculations presented in Ref.~\cite{Aslan2023}, the dependence on $\mu$ exactly cancels between the W- and Y-terms if the coupling is taken constant. In this specific case, the value of $\mu$ is irrelevant. However, we know that for general cases, a good practice is to set $\mu\sim Q$. Then, the $k_\T$ associated to the PDF is kept small, i.e., smaller than the hard scale, and large $k_\T$ are managed by the hard coefficient. It can be seen clearly from the W-term-like diagram contributing to the Y-term (i.e., the first diagram of Fig.~\ref{exact} with masses set to zero and with a convolution with the trivial zeroth order PDF) and its subtraction term 
\begin{equation}
    a_\lambda(1-x_{\text{bj}})\left(\int_0^{\hat{k}_m^2(x_{\text{bj}})}dk_\T^2 \frac{1}{\kappa(x_{\text{bj}}) k_\T^2}-\int_0^{\mu^2} dk_\T^2 \frac{1}{k_\T^2}\right), \label{ytfirst}
\end{equation}
with 
\begin{equation}
    \hat{k}_m^2(\xi)=\frac{(1-\xi)Q^2}{4\xi}, \label{limitkt2}
\end{equation}

\begin{equation}
    \kappa(\xi)=\sqrt{1-\frac{k_\T^2}{\hat{k}_m^2(\xi)}},
\end{equation}
and
\begin{equation}
    a_{\lambda}=\frac{\lambda^2}{16\pi^2}.
\end{equation}
For $\hat{k}_m^2\gg \mu^2\simeq Q^2$, which is the case at small $x_{\text{bj}}$, $\kappa \sim 1$ in the region $k_\T<\mu$. In this case, Eq.~(\ref{ytfirst}) approximately reduces to its first term integrated from $\mu^2$ to $\hat{k}_m^2$. We see that the W-term makes the integration from $0$ to $\mu^2$, explicitly,
\begin{equation}
    \mathrm{T}_{\mathrm{small}} F_2 = a_\lambda(1-x_{\text{bj}})x_{\text{bj}}\int_0^{\mu^2}dk_\T^2 \frac{k_\T^2+\chi(x_{\text{bj}})^2}{[k_\T^2+\Delta(x_{\text{bj}})^2]^2}, \label{expw}
\end{equation}
with $\Delta(x_{\text{bj}})$ and $\chi(x_{\text{bj}})$ two functions of small masses (whose precise definition is irrelevant here),
while the Y-term takes care of the remaining phase space.\\

One could have the idea to ignore Eq.~(\ref{ytfirst}) (and, more generally, the Y-term) and simply integrate the W-term up to $\hat{k}_m^2$. It is, in fact, what does the $k_t$-factorization if $\kmax = \hat{k}_m^2$. In the simple case of DIS in the Yukawa theory at order $\mathcal{O}(\lambda^2)$, it would not be necessarily a bad idea. However, we should question the validity of an integration above $Q^2$ in the general case.

\section{$k_{\text{\footnotesize max}}$ is the factorization scale \label{seckmax}}
In presenting Eqs.~(\ref{resumfa}) and (\ref{UCN}) for the resummation factor and the relation between collinear and unintegrated PDFs, we made a simplification, and it is necessary to revisit this point. It is instructive to consider the case where the factorization scale $\mu_F$ at which the resummation factor is factorized is not necessarily the same as the scale $\mu$ used to renormalized the collinear PDFs. Then we have
\begin{equation}
    C_N(k_\T,\alpha_s;\mu_F)=\gamma_N(\alpha_s)\frac{1}{k_\T^2}\left(\frac{k_\T^2}{\mu_F^2}\right)^{\gamma_N(\alpha_s)}, \label{resumfa2}
\end{equation}
\begin{equation}
    F_N(k_\T;\mu,\mu_F)= f_N(\mu)C_N(k_\T,\alpha_s;\mu_F), \label{fnmuf}
\end{equation}
and
\begin{equation}
    \int_0^{\mu_F^2}dk_\T^2 F_N(k_\T;\mu,\mu_F)= f_N(\mu)\int_0^{\mu_F^2}dk_\T^2 C_N(k_\T,\alpha_s;\mu_F) = f_N(\mu)\label{UCN2}.
\end{equation}
Of course, this equations reduces to Eq.~(\ref{UCN}) if $\mu_F=\mu$, which is probably the best choice.
It is also clear that $\kmax=\mu_F^2$. In particular, in the presentation of the $k_t$-factorization given in Ref.~\cite{Collins1991}, $\mu_F$ appears in the resummation factor because the contribution $k_\T<\mu_F$ has been subtracted from the impact factor\footnote{In Ref~\cite{Collins1991}, the impact factor is the off-shell cross section with the large $\hat{s}$ behavior removed.} in order to define a collinear safe hard coefficient (But Ref.~\cite{Collins1991} works with $\mu_F=\mu$). Then, $k_\T>\mu_F$ belong to the hard coefficient, and $k_\T<\mu_F$ to the UPDFs. Applying the UPDFs of Eq.~(\ref{fnmuf}) to DIS with the on-shell coefficient, we find that the result does not depend on the factorization scale, as it should. Explicitly, for $\mu_F=Q$ and $\mu_F=\sqrt{s}$, where $\sqrt{s}$ stands for any scale much larger than $Q$, we have
\begin{align}
    \frac{F_2}{x_{\text{bj}}}&= \text{TM}^{-1} \left\{ f_N(\mu)\int_0^{Q^2}dk_\T^2  C_N(k_\T,\alpha_s;Q)\right \}\nonumber \\
    &= \text{TM}^{-1} \left\{ f_N(\mu)\int_0^{s}dk_\T^2  C_N(k_\T,\alpha_s;\sqrt{s})\right \}=f(x_{\text{bj}};\mu), \label{muft}
\end{align}
where $\text{TM}^{-1}$ stands for the inverse Mellin transform.
The overestimation discussed earlier comes precisely from using UPDFs obeying Eq.~(\ref{UCN}), i.e., for $\mu_F=\mu \sim Q$, with the transverse-momentum of the $k_t$-factorization formula integrated up to $s$, i.e., $\kmax=\mu_F^2=s$. In other word, there is an inconsistent choice of the factorization scale. In our example for the structure function, it would lead
\begin{align}
    \frac{F_2}{x_{\text{bj}}}&= \text{TM}^{-1} \left\{ f_N(\mu)\int_0^{s}dk_\T^2  C_N(k_\T,\alpha_s;\mu)\right \}\\
    &=f(x_{\text{bj}};\mu)+\text{TM}^{-1} \left\{ f_N(\mu)\int_{\mu^2}^{s}dk_\T^2  C_N(k_\T,\alpha_s;\mu)\right \},
\end{align}
giving indeed an overestimation of the correct result. This discussion solves the issue mentioned in Refs.~\cite{Guiot2020a,Guiot2023} about the mismatch between Eq.~(\ref{ktfact}), with a transverse-momentum integration above $\mu$, and UPDFs built from Eq.~(\ref{UCN}), constrained only in the range $k_\T\in [0,\mu]$. From Eq.~(\ref{UCN2}) and $\kmax=\mu_F^2$, we see that the transverse-momentum integration in the $k_t$-factorization formula does not extend to a region where the UPDFs are not constrained.\\

From the point of view of the consistent treatment of the factorization scale, the two possibilities given in Eq.~(\ref{muft}) are equally acceptable. But the first one, $\mu_F=\mu\sim Q$, should be preferred because the second implies a kinematics with large $k_\T$, which would be better described by the hard coefficient, and leads to the complicated relation (\ref{UCN2}) with $\mu_F^2=s$. Note, however, that approximating $s \sim \infty$, these UPDFs would obey
\begin{equation}
    \int_0^{\sim \infty}dk_\T^2 F(x,k_\T;\mu)=f(x;\mu).\label{relinf2}
\end{equation}
Ref.~\cite{Guiot2023} showed that, while the KMRW UPDFs [which obey Eq.~(\ref{UCN}) with a factor $x$, see Sec.~\ref{secKMR} for more details] used with the \KaTie\ event generator overestimate the D-meson cross section, a modified version of these UPDFs obeying Eq.~(\ref{relinf2}) (with a factor $x$ in the r.h.s.) don't. As already discussed in Sec.~\ref{secmotiv}, \KaTie\ integrates the transverse-momentum significantly above the hard scale for this observable,\footnote{However, the initial transverse momenta can be restricted in the \KaTie\ input file, then changing the default values.} and the good agreement obtained with the modified KMRW UPDFs is the consequence of the consistent choice $\mu_F \gg Q$. On the opposite, Ref.~\cite{Guiot2021} obtained good results for the same observable, using the usual KMRW UPDFs, and cutting the integral at $k_\T=Q$ in \KaTie, corresponding to the choice $\mu_F=\mu=Q$.\\

We will see that, for various reasons, identifying situations where a substantial disagreement with the exact result arises is not necessarily straightforward. One of these reasons is a special case of UPDFs, presented in the next section along with the KMRW UPDFs. The latter will be employed to exemplify our observations through explicit calculations.

\section{Two kinds of UPDFs \label{secKMR}}
In the Yukawa theory, finite collinear PDFs can be defined in two ways. Either by a cutoff on the transverse-momentum integral of TMD PDFs
\begin{equation}
    f(x;\mu)=\int_0^{\mu^2}d^2 k_\T f(x, k_\T;\mu), \label{reluptmd}
\end{equation}
or by applying the Feynman rules for collinear PDFs and renormalized the UV divergences. At least at order $\lambda^2$, the Feynman rules for collinear PDFs are equivalent to integrating the TMD PDFs to infinity. Then, we have
\begin{equation}
    f(x;\mu) = \int_0^{\infty}d^2 k_\T f(x, k_\T;\mu) + \overline{\text{MS}}\text{ C.T.}, \label{fren}
\end{equation}
with the second term corresponding to the modified minimal-subtraction-scheme counterterm. \\

We already mentioned that UPDFs such as the KMRW obey Eq.~(\ref{reluptmd}), with a factor $x$ to the left
\begin{equation}
   x f(x;\mu)=\int_0^{\mu^2}d^2 k_\T F(x, k_\T;\mu). \label{relup}
\end{equation}
However, this factor comes from the fact that these UPDFs are applied to the cross section differential in the rapidity $y$, while TMD factorization is used for the cross section differential in $x$. The change of variable between these two quantities involves a factor $x$. Then, we should compare the TMD PDFs to the UPDFs divided by $x$. Note also that we generally have an integration on $dk_\T^2$ rather than on $d^2k_\T$ for UPDFs. In this paper, we will work with Eq.~(\ref{relup}) and add a factor $1/\pi$ in the definitions of UPDFs found in the literature.\\

Another kind of UPDFs are related to collinear PDFs by
\begin{equation}
     xf(x;\mu)=\int_0^{\infty}d^2 k_\T \mathcal{F}(x, k_\T;\mu). \label{relinf}
\end{equation}
Here and in the following, we use the notation $F$ and $\mathcal{F}$ for UPDFs obeying Eqs.~(\ref{relup}) and (\ref{relinf}), respectively. We have already met an equation similar to Eq.~(\ref{relinf}) earlier, but it was an approximation, and the nature of the UPDFs $\mathcal{F}$ discussed in the following is different. One could ask why Eq.~(\ref{relinf}) is not divergent, since we may expect the unintegrated PDFs to scale like $1/k_\T^2$ at large $k_\T$, both in the Yukawa theory and in QCD. But Eq.~(\ref{fren}) suggest that we can define
\begin{equation}
    \mathcal{F}/x=F/x + \overline{\text{MS}}\text{ c.t.}(\mu), \label{Fpct}
\end{equation}
with $\overline{\text{MS}}\text{ c.t.}$ the integrand of the counterterm
\begin{equation}
   (2\pi\mu)^{2\epsilon} \int d^{2-2\epsilon}k_\T \, \overline{\text{MS}}\text{ c.t.}(\mu) = \overline{\text{MS}}\text{ C.T.}
\end{equation}
The counterterm $\overline{\text{MS}}\text{ c.t.}$ depends on $\mu$ (but not on $\epsilon$).
We will give an explicit example in the Yukawa theory at the end of this section.\\

It is interesting to compute the KMRW UPDF in the Yukawa theory and compare them with the TMD PDF given in Ref.~\cite{Aslan2023}:
\begin{equation}
    f(x,k_\T;\mu)=\frac{a_\lambda}{\pi}(1-x)\frac{k_\T^2+\chi(x_{\text{bj}})^2}{[k_\T^2+\Delta(x_{\text{bj}})^2]^2}, \label{tmdpdf}
\end{equation}
with 
\begin{equation}
    \Delta(x)^2=xm_s^2+(1-x)m_q^2-x(1-x)m_p^2, \quad \chi(x)^2=(m_q+xm_p)^2.
\end{equation}
We start with the usual definitions \cite{Kimber2001,Watt2003}
\begin{align}
F_q(x,k_t;\mu)&=\frac{x}{\pi}\frac{\partial}{\partial k_t^2}\left[T_q(k_t,\mu)f_q(x,k_t)\right], \quad k_t \geq \mu_0\label{def1}\\
F_q(x,k_t;\mu^2)&=\frac{x}{\mu_0^2\pi} T_q(\mu_0,\mu)f_q(x,\mu_0), \quad k_t<\mu_0,\label{conkt}
\end{align}
and we choose $\mu_0=m_s=1$ GeV by simplicity. $T_q$ is the Sudakov factor, and it is enough to say it is given by an exponential whose exponent is proportional to $a_\lambda$. At order $\mathcal{O}(a_\lambda)$, we should then consider the products $T_q^{(1)}f_q^{(0)}$ and  $T_q^{(0)}f_q^{(1)}$. However, since $f_q^{(0)}\propto \delta(1-x)$, the first product is irrelevant for our study. The zeroth order of the Sudakov factor is 1, and we are left with the calculation of
\begin{align}
    F_q(x,k_t;\mu)/x&=\frac{1}{\pi}\frac{\partial}{\partial k_t^2}\left[f_{q/p}^{(1)}(x,k_t)\right]\\
    &= \frac{a_\lambda}{\pi k_\T^2}(1-x), \quad k_\T>\mu_0\label{kmrpdf2}
\end{align}
where we used the expression of $f_{q/p}^{(1)}$ given by Eq.~(23) of Ref.~\cite{Aslan2023}. Eq.~(\ref{kmrpdf2}) is the same as Eq.~(\ref{tmdpdf}) with all the masses set to zero. There is a dependence on $\mu$ through the coupling constant, but we will ignore its running. Similarly, we find for $k_\T<\mu_0$
\begin{align}
    F_q(x,k_t;\mu^2)/x&=\frac{1}{\mu_0^2 \pi} f_{q/p}^{(1)}(x,\mu_0)\\
    &=\frac{1}{\mu_0^2 \pi}a_\lambda (1-x)\left(\frac{\chi(x)^2}{\Delta(x)^2}+\ln\left(\frac{\mu_0^2}{\Delta(x)^2}\right)-1\right), \quad k_\T<\mu_0 \label{kmry2}
\end{align}
It is possible to show that
\begin{equation}
    (2\pi\mu)^{2\epsilon}\int_0^\infty d^{2-2\epsilon}k_\T F_q(x,k_\T;\mu)/x + \overline{\text{MS}}\text{ C.T.} \label{infct}
\end{equation}
gives exactly the collinear PDFs of Ref.~\cite{Aslan2023}.\footnote{It was not guaranteed because the cutoff and renormalization methods are only equivalent up to $\mathcal{O}(m/\mu)$ corrections, and the KMRW UPDFs are defined from Eq.~(\ref{relup}).} To do so, we start by
\begin{equation}
    A=\int_0^{\mu_0^2}d^2k_\T F_q(x,k_\T;\mu)/x = a_\lambda (1-x)\left(\frac{\chi(x)^2}{\Delta(x)^2}+\ln\left(\frac{\mu_0^2}{\Delta(x)^2}\right)-1\right).
\end{equation}
Then, we split the second integral, for $k_\T>\mu_0$, into two parts
\begin{equation}
    (2\pi\mu)^{2\epsilon} \int_{\mu_0^2}^\infty d^{2-2\epsilon}k_\T F_q =  (2\pi\mu)^{2\epsilon} \int_{0}^\infty d^{2-2\epsilon}k_\T F_q - \int_{0}^{\mu_0} d^2k_\T F_q = x(B - C). \label{termBC}
\end{equation}
We set $\epsilon=0$ in the term $C$ which is not UV divergent.
To regularize the IR divergence, we modify the denominator of Eq.~(\ref{kmrpdf2}) to
\begin{equation}
    \frac{1}{k_\T^2}\to \frac{1}{k_\T^2+\delta}.
\end{equation}
The regulator will cancel when we add the $B$ and $C$ terms. Using the relation 
\begin{equation}
    \int d^d\bm{k}f(\bm{k})=\frac{\pi^{d/2}}{\Gamma\left(\frac{d}{2} \right)}\int_0^{\infty}dk^2(k^2)^{d/2-1}f(k^2),
\end{equation}
with $d=2-2\epsilon$, we find
\begin{equation}
    B=(2\pi\mu)^{2\epsilon}a_\lambda (1-x)(\pi \delta)^{-\epsilon}\Gamma(\epsilon),
\end{equation}
and
\begin{equation}
    C=a_\lambda (1-x) \ln\left(\frac{\mu_0^2+\delta}{\delta}\right).
\end{equation}
Using the expression for $\Gamma(\epsilon)$ and performing the expansion in $\epsilon$ we find
\begin{equation}
    B-C=a_\lambda(1-x)\left(\frac{1}{\epsilon}-\gamma_E +\ln (4\pi\mu^2) -\ln(\mu_0^2+\delta) +\mathcal{O}(\epsilon)\right)
\end{equation}
Finally, adding $A$ and the $\overline{\text{MS}}$ counterterm, and taking $\epsilon, \delta\to 0$, Eq.~(\ref{infct}) gives
\begin{equation}
     a_\lambda(1-x)\left[\frac{\chi(x)^2}{\Delta(x)^2}+\ln\frac{\mu^2}{\Delta(x)^2} -1\right] = f_{q/p}^{(1)}(x,\mu),
\end{equation}
which is exactly Eq.~(23) of Ref.~\cite{Aslan2023}.\\

We close this section by performing explicitly the calculation where the integrand of the $\overline{\text{MS}}$ countertern is included in the definition of UPDFs, see Eq.~(\ref{Fpct}), leading to UPDFs obeying Eq.~(\ref{relinf}). With our toy calculation, we want to show that relation (\ref{relinf}) is not necessarily surprising, and illustrate a general property of such parton densities. A direct calculation shows that
\begin{equation}
    \overline{\text{MS}}\text{ C.T.}=a_\lambda(1-x)\left(\frac{1}{\epsilon}-\gamma_E+\ln 4\pi \right)+\mathcal{O}(\epsilon)=(2\pi\mu)^\epsilon\int d^{2-2\epsilon}k_\T\frac{a_\lambda}{\pi}\frac{1-x}{(k_\T^2+\mu^2)}.\label{ctint}
\end{equation}
Then we find
\begin{equation}
    \overline{\text{MS}}\text{ c.t.}(\mu) = \frac{a_\lambda}{\pi}\frac{1-x}{(k_\T^2+\mu^2)},\label{intct}
\end{equation}
and
\begin{eqnarray}
    \mathcal{F}(x,k_\T;\mu)/x = \frac{a_\lambda}{\pi}(1-x)\frac{\mu^2}{k_\T^2(k_\T^2+\mu^2)}, \label{kmrupdf1}
\end{eqnarray}
This equation is obtained by summing Eqs.~(\ref{intct}) and (\ref{kmrpdf2}).
If we want to stick more strictly to the KMRW ``style'' (and also have the correct mass terms), we should replace $\mu^2$ by $\mu^2-\mu_0^2$ in the denominator of Eq.~(\ref{kmrupdf1}), leading to
\begin{align}
     \mathcal{F}(x,k_\T;\mu)/x &=\frac{a_\lambda}{\pi}(1-x)\frac{\mu^2}{k_\T^2(k_\T^2+\mu^2-\mu_0^2)}, \quad k_\T > \mu_0 \label{renkmr2}\\
     \mathcal{F}(x,k_\T;\mu)/x &=\frac{1}{\mu_0^2 \pi}a_\lambda (1-x)\left(\frac{\chi(x)^2}{\Delta(x)^2}+\ln\left(\frac{\mu_0^2}{\Delta(x)^2}\right)-1\right), \quad k_\T < \mu_0. \label{renkmr1}
\end{align}
By construction, this function obeys Eq.~(\ref{relinf}).\\

\begin{figure}[!h]
\begin{center}
 \includegraphics[width=21pc]{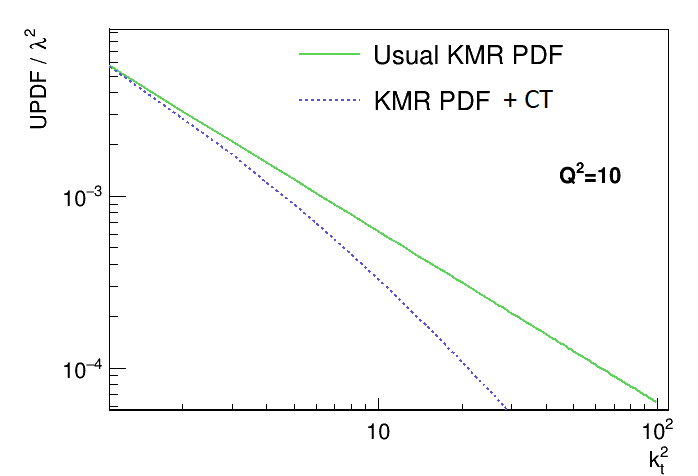}
\end{center}
\caption{\label{fcompkmr} Comparison of standard KMR UPDF given by Eqs. (\ref{kmrpdf2}) and (\ref{kmry2}) with the version having the counterterm included, Eqs. (\ref{renkmr1}) and (\ref{renkmr2}). We used $x=0.01$ and $\mu=Q$.}
\end{figure}

In Fig.~\ref{fcompkmr}, we compare the usual KMRW UPDF with the one having the counterterm included. We see that the numerical values of both functions are close for $k_\T<Q$. On the opposite, the blue dotted line obtained with Eqs.~(\ref{renkmr2})-(\ref{renkmr1}) is significantly below the green line in the region $k_\T>Q$. 
This feature is typical of UPDFs obeying Eq.~(\ref{relinf}) such as PB \cite{Hautmann2017,Martinez2019} and CCFM-K \cite{Gawron2003} parton densities. 
However, the former go to zero quite faster than the dotted blue line of Fig.~\ref{fcompkmr}. We should also remember that the results shown in Fig.~\ref{fcompkmr} have been obtained in the Yukawa theory and not in QCD. But we can imagine that an equation similar to Eq.~(\ref{Fpct}) is also true in QCD (for the UPDFs) since nothing forbid to decompose $\mathcal{F}$ into a function $F$ scaling like $1/k_\T^2$ at large transverse momentum and a counterterm, which may differ from the $\overline{\text{MS}}$ counterterm. Such a claim would require a deeper study, which is not the goal of this work. Here, we wanted to illustrate that the integration of $\mathcal{F}$ to infinity can give a finite quantity. Based on our particular example in the Yukawa theory, general considerations, and known examples (such as the PB UPDFs) in QCD, we also consider likely that, in QCD, any function $\mathcal{F}$ have the behavior discussed above, i.e., a negligible value for $k_\T>\mu$.\\

\begin{figure}[!h]
\begin{center}
 \includegraphics[width=24pc]{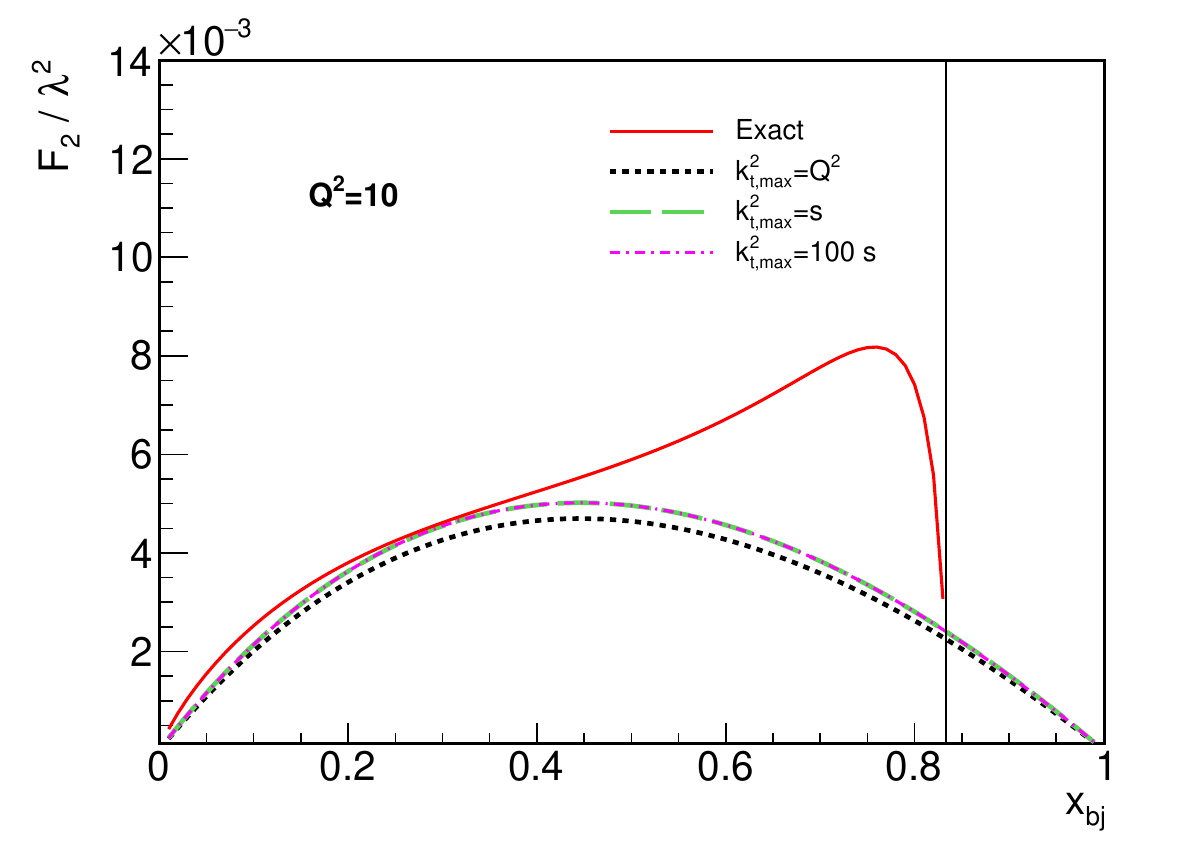}
\end{center}
\caption{\label{f2kmrRen}  Comparison of the exact $\mathcal{O}(a_\lambda)$ result with $k_t$-factorization calculations for the $F_2$ structure function. We used the UPDFs defined in Eqs.~(\ref{renkmr1}) and (\ref{renkmr2}). The dotted, dashed, and dashed-dotted lines correspond to different values of $\kmax$. The vertical line indicates the kinematical limit for the exact case.}
\end{figure}

In Fig.~\ref{f2kmrRen}, we compare the exact structure function to the $k_t$-factorization result (using the on-shell coefficient) obtained with the UPDFs defined in Eqs.~(\ref{renkmr1}) and (\ref{renkmr2}). 
The transverse momentum component has been integrated up to $Q^2$ (black dotted line), $\sqrt{s}$ (green dashed line), and $s\times 10^2$ (pink dashed-dotted line). In principle, the integration range could extend to infinity, given that the counterterm embedded in the definition of $\mathcal{F}$ cuts the large $k_\T$ contribution. Indeed, we can see that changing the integration limit from $s$ to $s\times 10^2$ has no effect. In practice, it is due to the negligible value of this UPDF in the region $k_\T \gg \mu$. Even if the black line is slightly below the two others (using the UPDFs with a counterterm and cutting the integral at $Q^2$ results in an oversubtraction) the comparison with the exact result is of similar quality, in particular at small $x_{\text{bj}}$. Then, for UPDFs obeying Eq.~(\ref{relinf}), the precise definition of $\kmax$ does not matter, and any $\kmax > Q^2$ will do the job.

\newpage
\section{Discussion and conclusion \label{secQCD}}

The most important observation is that the $k_t$-factorization must integrate the transverse momentum up to the factorization scale, i.e., $\kmax = \mu_F^2$, and the latter is not necessarily the same than the scale $\mu$ of the collinear PDFs. We showed that an inconsistent choice of the factorization scale, with $\mu_F=\mu\sim Q$ in the resummation factor and $\mu_F \gg Q$ in Eq.~(\ref{ktfact}), leads potentially to an overestimation of the cross section. In sec.~\ref{seckmax}, we perfomed explicit calculations in QCD using the UPDFs given in the seminal papers on $k_t$-factorization. But it is clear that the conclusion extends to all UPDFs obeying Eqs.~(\ref{UCN}), (\ref{UCN2}), or (\ref{relup}).

Nonetheless, even when the factorization scale lacks consistency, with the situation just described above, specific kinematics can hide the issue. This last point can be illustrated as follows.  We have seen that for proton-proton collisions, the transverse momentum is sometimes integrated significantly above $Q$, which is the case, in particular, with the event generator \KaTie. In those situations, UPDFs obeying Eq.~(\ref{relup}) give potentially an overestimation of the cross section. We gave an explicit example for the transverse-momentum distribution of a gluon, see Fig.~\ref{ggtogg}, where the result obtained with the KMRW UPDFs (top red line) is significantly above the green and blue lines. 
Now, if we keep playing with the $gg\to gg$ process but require the transverse momentum of the two outgoing gluons to lie in the range $[200,210]$ GeV, the PB and KMRW functions give a total cross section of 2.7 and 3.1 nb, respectively. Clearly, the nearly symmetric configuration favors small $k_\T<\mu \sim 200$ GeV, making the integration for $k_\T >\mu$ irrelevant. Then, the two sets of UPDFs lead to similar results.\\

The previous sections suggested three possible implementations of the $k_t$-factorization. Keeping in mind that $\kmax=\mu_F^2$ and $\mu \sim Q$, we can use
\begin{enumerate}
    \item UPDFs obeying Eq.~(\ref{relinf}) with $\kmax \gg \mu^2$ 
    \item UPDFs obeying Eq.~(\ref{relup}) [or Eq.~(\ref{UCN})] with $\kmax=\mu^2$
    \item UPDFs obeying Eq.~(\ref{UCN2}) with $\kmax \gg  \mu^2$. It reduces to option 2. for $\kmax=\mu^2$. 
\end{enumerate}
The first two options are close because we saw that the UPDFs of Eq.~(\ref{relinf}) give a negligible contribution for $k_\T>\mu$. This point was discussed in Sec.~\ref{secQCD} for the Yukawa theory, and we argued it is likely to be the case also in QCD. At least, known examples such as the PB UPDFs show this behavior. Then, this type of parton densities can be used with the event generator \KaTie\ without further consideration. On the opposite, UPDFs obeying Eq.~(\ref{relup}), such as the KMRW UPDFs, can be employed with \KaTie\ only if the transverse-momentum integration is cut at $k_\T=\mu$ (such cut can be implemented in the \KaTie\ input file). The third option seems unnecessarily complicated, which may reflect an artificial separation of $\mu_F$ and $\mu$.\\

Finally, another possibility, not satisfying from the theoretical point of view, but maybe efficient at leading order, is to use the second proposition above, with $\mu=\hat{k}_m$ instead of $Q$, where $\hat{k}_m$ is the kinematical limit. One reason this choice is unsatisfactory is that $\hat{k}_m$ is generally a function of the convolution variable $\xi$, see Eq.~(\ref{limitkt2}) (or $\hat{k}_m=\hat{s}=\xi_1\xi_2 s$ in pp collisions). It is well-established that if $\mu$ depends on this variable, higher-order corrections do not reduce the scale uncertainty \cite{Collins1989}. Then, a better solution is to keep $\mu \sim Q$ and include higher-order contributions. This task being quite complicated for off-shell quantities, a possibility is matching the LO high-energy cross section to collinear factorization
\begin{equation}
    \sigma = \sigma_{\text{HEF}} -\text{subtraction term} +\sigma_{\text{FO}}, \label{matchHEF}
\end{equation}
see, for instance, Refs.~\cite{Ball2001,Lansberg2022}. Going beyond leading order is probably unavoidable for observables such as the transverse-momentum distribution of $J/\psi$ particles. At LO, the $J/\psi$ transverse momentum is entirely determined by the UPDFs, except for the fragmentation contribution. Fig. 3 of Ref.~\cite{Maciula2019} demonstrates that KMRW and CCFM-based UPDFs lead to incompatible results for this observable, with the former showing a much better agreement with data. While one could conclude on the superiority of the KMRW formalism, in our opinion, it underlines the lack of reliability of LO calculations for this observable. Indeed, as reported in \cite{Guiot2023a}, we found that NLO calculations using CCFM-like UPDFs describe data well, and that the large-$p_t$ $J/\psi$ come from higher-order processes which cannot all be accounted for by UPDFs and initial-state radiations. Again, a natural choice in Eq.~(\ref{matchHEF}) is to describe small and large $p_t$ with $\sigma_{\text{HEF}}$ and $\sigma_{\text{FO}}$, respectively. In other words, to set $\kmax = \mu^2$ in $\sigma_{\text{HEF}}$.\\

We close this discussion with a word on the estimation of scale uncertainties. Even if we set $\mu_F=\mu$ for central values of the cross section, these two scales can be varied independently. This situation is similar to the case $\mu_R=\mu$, where $\mu_R$ is the renormalization scale associated with the strong coupling constant. Explicitly, we could compute the cross section for $\mu_F,\, \mu=Q$, $Q/2$, and $2Q$, leading to nine results that can be used to estimate the uncertainty. However, we should remember that the factorization scale $\mu_F$ appears both as an upper limit of integration and in the UPDFs. Consequently, varying independently the two scales is possible only if the UPDFs retain their dependence on both $\mu$ and $\mu_F$. If the dependence on $\mu_F$ is partially lost (because the UPDFs are provided with $\mu_F=\mu$), the procedure just described above is not applicable. Then, we should set $k_{\text{\footnotesize max}}=\mu_F=\mu$ and obtain three estimates of the cross section (including the central value) after the common variation by a factor of 2. In our example for DIS with an on-shell coefficient, see Eq.~(\ref{muft}), the difference between these two cases is unimportant. The dependence on $\mu_F$ exactly cancels in the integral, and the final result is $f(x_{\text{bj}};\mu=Q)$ with the uncertainty set by $f(x_{\text{bj}};2Q)$ and $f(x_{\text{bj}};Q/2)$, no matter if both scales vary independently or not. It would not be the case if the hard coefficient were $k_\T$ dependent.

\section*{Acknowledgments}
BG acknowledge support from ANID PIA/APOYO AFB230003. AvH is supported by grant no.\ 2019/35/B/ST2/03531 of the Polish National Science Centre. We are grateful to Francesco Hautmann for his comments on the manuscript.

\bibliographystyle{BibFiles/t1}
\bibliography{BibFiles/GeneralBib}
\end{document}

%% file: Packages/myTikz.tex
\usepackage{tikz}

\usetikzlibrary{decorations.markings}
\usetikzlibrary{decorations.pathmorphing}

\tikzset{
    vector/.style={decorate, decoration={snake}, draw},
    fermion/.style={draw=black, postaction={decorate},
        decoration={markings,mark=at position .55 with {\arrow[scale=1.,>=stealth]{>}}}},
    fermionbar/.style={draw=black, postaction={decorate},
        decoration={markings,mark=at position .58 with {\arrow[scale=1.,>=stealth]{<}}}},
    fermionline/.style={draw=black},
    gluon/.style={decorate, draw=black,
        decoration={coil,amplitude=4pt, segment length=5pt}},
    scalar/.style={dashed, draw=black, postaction={decorate},
        decoration={markings,mark=at position .55 with {\arrow[scale=1.,>=stealth]{>}}}},
    scalarbar/.style={dashed, draw=black, postaction={decorate},
        decoration={markings,mark=at position .58 with {\arrow[scale=1.,>=stealth]{<}}}},
    scalarline/.style={dashed,draw=black},
    fmassin/.style={draw=black, postaction={decorate},
    decoration={markings, mark=at position .75 with {\arrow[scale=1.,>=stealth]{>}}, mark=at position .30 with {\arrow[scale=1.,>=stealth]{<}}}},
}

%% file: Packages/myHypref.tex
\usepackage[colorlinks=true,linkcolor=redLinks,citecolor=greenLinks,urlcolor=redLinks, pdfborder={0 0 1}]{hyperref}
\definecolor{greenLinks}{rgb}{0, 0.6, 0} 
\definecolor{blueLinks}{rgb}{0, 0, 0.6}
\definecolor{redLinks}{rgb}{0.6, 0, 0}
\definecolor{blackLinks}{rgb}{0.77,0.12,0.23} 
\definecolor{eprintLinks}{rgb}{0.4, 0.4, 0.4}
\definecolor{journalLinks}{rgb}{0.6, 0, 0}